\documentclass[aps,prd,english,superscriptaddress, longbibliography,11pt,notitlepage]{revtex4}
\usepackage[colorlinks=true,  pdfstartview=FitV,
linkcolor=blue, citecolor=blue, urlcolor=blue]{hyperref}

\usepackage{amsmath}
\usepackage{amssymb}
\usepackage{graphicx}
\usepackage{hhline}
\usepackage{mwe}
\usepackage{amsbsy}
\usepackage{textcomp}
\usepackage{commath}
\usepackage{subfig}
\usepackage{slashed}
\usepackage{natbib}

\usepackage{stackengine}
\stackMath

\makeatletter
\usepackage{babel}
\numberwithin{equation}{section}
\begin{document}
\immediate\write16{<<WARNING: LINEDRAW macros work with emTeX-dvivers
                    and other drivers supporting emTeX \special's
                    (dviscr, dvihplj, dvidot, dvips, dviwin, etc.) >>}

\title{ Abelian Chern-Simons vortices in the presence of magnetic impurities }
%\preprint{KA-TP-??-2017}
%\date{23-03-2022}
\author{D. Bazeia}
\affiliation{Departamento de Física, Universidade Federal da Paraíba, João Pessoa - PB - 58051-970, Brazil}
\author{J. G. F. Campos}
\affiliation{Departamento de Física, Universidade Federal de Pernambuco, Recife - PE - 50670-901, Brazil}
\author{A. Mohammadi}
\affiliation{Departamento de Física, Universidade Federal de Pernambuco, Recife - PE - 50670-901, Brazil}

\begin{abstract}

This work deals with Abelian Chern-Simons vortices interacting with magnetic impurities. We compute static solutions with winding numbers zero and one. Then, we develop a numerical algorithm to simulate their collisions. Collisions between a vortex with winding number two and a magnetic impurity are also performed. All scattering results are interpreted in terms of the moduli space approximation and compared with the Abelian Maxwell-Higgs model.

\end{abstract}

\maketitle

%%%%%%%%%%%%%%%%%%%%%%%%%%%%%%
\section{Introduction}
\label{intro}

Vortices represent topological solutions within field theories where an order parameter winds around its center once or multiple times. In two spatial dimensions, the vortex center is a single point, a string in three dimensions, and a $(D-2)$-dimensional surface in $D>3$ dimensions. Such solutions are ubiquitous across various domains including fluids \cite{Landau:2013}, superfluids \cite{Salomaa:1987}, superconductors \cite{Ginzburg:1950, Abrikosov:1957}, cosmic strings \cite{Vilenkin:1994}, particle physics \cite{James:1994}, string theory \cite{Tong:2002}, and numerous other contexts.

A canonical description of relativistic vortices is given by the Abelian Maxwell-Higgs (AMH) model \cite{Manton:2004}. This model, renowned for its demonstration of superconductivity, serves as a prototype for describing spontaneous symmetry-breaking phenomena \cite{Higgs:1964}. Additionally, the AMH model describes cosmic strings when coupled to gravity \cite{Vilenkin:1994}. Another significant relativistic model featuring gauged vortices is the Abelian Chern-Simons (ACS) model \cite{Hong:1990, Jackiw:1990a, Jackiw:1990b}. It consists of the Chern-Simons gauge field coupled with a Higgs scalar field, with the Lagrangian characterizing a topological field theory in (2+1)-dimensions, thus revealing phenomena such as the fractional quantum Hall effect and anyons \cite{Atland:2010}.

Generally, computing the vortex profile within the AMH model demands numerical techniques, both in the general case \cite{Harden:1963, Plohr:1981, Berger:1989} and in the Bogomol'nyi–Prasad–Sommerfield (BPS) limit \cite{Jaffe:1980}. The same holds for ACS vortices \cite{Jackiw:1990a,Jackiw:1990b}. However, analytical vortex solutions can be constructed for certain modified theories (see for instance \cite{Bazeia:2018c}). Intriguingly, the internal structure of vortices can be manipulated in various ways by modifying the Lagrangian \cite{Bazeia:2010, Bazeia:2018, Bazeia:2018b, Bazeia:2019, Andrade:2020}.

The scattering behavior of AMH vortices was initially investigated in Refs.~\cite{Ruback:1988, Shellard:1988}. Through both numerical simulations and analytical approaches, these studies revealed a remarkable finding that in a head-on collision, two vortices scatter at a precise 90-degree angle. The authors attributed this unexpected result to the intricate geometry of the moduli space, which governs the dynamics of such solutions. In a more general scenario, the head-on collision of $n$ vortices scatters at a $\pi/n$ angle.

A key idea to compute the evolution of BPS solutions with small velocities is the concept of a moduli space \cite{manton:1982, manton:1985}. Such space is defined as the manifold which describes the set of BPS solutions. In many cases, the evolution of slowly moving BPS solutions can be described as a geodesic motion governed by the moduli space metric. A free fall is expected because BPS solutions are energetically equivalent and, consequently, potential interactions are absent. For the AMH vortex, a numerical scheme to compute the moduli space metric was proposed in Refs. \cite{Ruback:1988, samols:1992, Strachan:1992}. 

 The moduli space evolution of ACS vortices was obtained in Refs.~\cite{dziarmaga:1994, dziarmaga:1995}. The authors showed that the moduli space metric of ACS shares several features of the AMH, such as the $\pi/n$ scattering. However, significant disparities exist in the complete vortex dynamics between the two theories, mainly because the ACS vortices possess an electric charge and experience magnetic forces due to the magnetic fields of neighboring vortices. Despite the moduli space dynamics of ACS vortices being computed a few years after those of AMH vortices, actual scattering simulations were conducted many years later \cite{Strilka:2012}. The scattering output reported indeed confirms the theoretical predictions.

A captivating modification of vortex dynamics involves the incorporation of magnetic impurities in superconducting materials. In the supersymmetric context, half the BPS property is guaranteed to be preserved by including a magnetic impurity in the AMH Lagrangian \cite{Hook:2013, Tong:2014}. Likewise, the same result holds for the ACS model \cite{Han:2016}. Those models are particularly fascinating because BPS solutions are energetically equivalent. Therefore, near BPS evolution can be well-approximated by moduli space dynamics. Moreover, they find application in the holography context \cite{ kachru:2010, benicasa:2012a, benicasa:2012b}.

In Refs. \cite{Cockburn:2017, Ashcroft:2020}, the authors explored the vortex profiles and the scattering of such systems in the bosonic sector. Specifically, they examined the AMH model with a Gaussian magnetic impurity. A similar analysis was performed in Ref.~\cite{Almeida:2022} for self-dual $CP(2)$ vortices. Interestingly, the negative energy density stemming from the impurity can be removed by treating it as a scalar field and adding its corresponding kinetic term to the Lagrangian \cite{Bazeia:2022}.

In the present work, we will study the vortex and vacuum configurations of the ACS model in the presence of BPS-preserving magnetic impurities. Subsequently, employing the obtained field profiles, we conduct simulations to scrutinize the scattering behavior of Chern-Simons vortices by these magnetic impurities. Two of our main goals are to interpret the scattering output regarding the moduli space dynamics and contrast our results with the behavior of AMH vortices investigated in Refs.~\cite{Cockburn:2017, Ashcroft:2020}. 

The paper is organized as follows. In section \ref{model}, we describe the Abelian Chern-Simons model incorporating a half-BPS preserving magnetic impurity. Section \ref{static} is devoted to computing the vacuum and vortex solutions in the presence of the magnetic impurity. In section \ref{scattering}, we delve into the moduli space dynamics and present a numerical algorithm devised to simulate the scattering phenomena between vortex solutions and magnetic impurities. Finally, the concluding section offers a summary of our findings and a discussion regarding their implications.

 \section{Abelian Chern-Simons Model}
 \label{model}
 
 Let us start with the following field theory in standard Minkowski spacetime in $(2+1)$ dimensions, with signature $(+,-,-)$
 \begin{equation}
     \mathcal{L}=\frac{\kappa}{4}\epsilon^{\alpha\beta\gamma}A_\alpha F_{\beta\gamma}+|D_\mu\phi|^2-V(|\phi|),
 \end{equation}
representing the so-called Abelian Chern-Simons model, where $\epsilon^{\alpha\beta\gamma}$ is the Levi-Civita tensor, $A_\mu$ is a $U(1)$ gauge field and $\phi$ is a complex scalar field. The field tensor and the covariant derivative are given by $F_{\mu\nu}=\partial_\mu A_\nu-\partial_\nu A_\mu$ and  $D_\mu\phi=\partial_\mu\phi+iA_\mu\phi$, respectively. Moreover, the scalar field potential is defined as
 \begin{equation}
     V(|\phi|)=\frac{1}{\kappa^2}|\phi|^2(\eta^2-|\phi|^2)^2.
 \end{equation}
 This form of $V(|\phi|)$ is often chosen to allow the Bogomol'nyi trick to be performed, as described below. The model leads to the following equations of motion for the fields
\begin{align}
    D_\mu D^\mu\phi=-\frac{\partial V}{\partial\phi^*},\\
    \frac{1}{2}\kappa\epsilon^{\mu\beta\gamma}F_{\beta\gamma}=J^\mu,
\end{align}
where the conserved matter current is given by
\begin{equation}
    J_\mu=i\left[\phi^* D_\mu\phi-\phi (D_\mu\phi)^*\right].
\end{equation}

The equations of motion can be simplified using the Bogomol'nyi trick as described in Refs.~\cite{Hong:1990,Jackiw:1990a,Jackiw:1990b}. The trick goes as follows. The Hamiltonian of the system is given by
\begin{equation}
    E=\int d^2r \left[|D_0\phi|^2+|D_i\phi|^2+V(\phi)\right].
\end{equation}
After using Gauss's law and performing some algebra, the energy of static solutions can be put in the following form
\begin{equation}
    E_{\text{static}}=\int d^2r \left[\left(\frac{\kappa}{2}\frac{B}{|\phi|}\pm\frac{|\phi|}{\kappa}(|\phi|^2-\eta^2)\right)^2+|D_1\phi\pm iD_2\phi|^2\pm \eta^2B\right],
\end{equation}
where $B=F_{21}$. Therefore, it is minimized if we set
\begin{align}
    D_1\phi\pm iD_2\phi=0,\\
    B=\pm\frac{2}{\kappa^2}|\phi|^2(\eta^2-|\phi|^2).
\end{align}
These are the BPS equations leading to the vortex solutions. Before solving the equations, let us perform an important modification. We will include a magnetic impurity that preserves half of the system's BPS structure. 

 \subsection{Magnetic Impurities}

To include a magnetic impurity that preserves half the BPS structure, the Lagrangian density is modified as follows
 \begin{equation}
     \mathcal{L}=\frac{\kappa}{4}\epsilon^{\alpha\beta\gamma}A_\alpha F_{\beta\gamma}+|D_\mu\phi|^2-\frac{1}{\kappa^2}|\phi|^2(\eta^2+\sigma(x)-|\phi|^2)^2+\sigma(x)B.
 \end{equation}
 The term proportional to $B$ is the usual interaction between a magnetic moment $\sigma(x)$ and the magnetic field $B$. The scalar potential is then modified to maintain the BPS structure. The Bogomol'nyi trick goes almost unchanged as shown in Ref.~\cite{Han:2016}. Thus, we obtain the following first-order equations
\begin{align}
\label{eq:BPS1}
    D_1\phi+iD_2\phi=0\\
    B=\frac{2}{\kappa^2}|\phi|^2(\eta^2+\sigma(x)-|\phi|^2).
\label{eq:BPS2}
\end{align}
These equations lead to vortex solutions in the presence of magnetic impurities. Notice that only one sign choice remains because the other BPS solution is not preserved after adding the impurity. 

We will focus on a localized and spherically symmetric impurity of the Gaussian type
\begin{equation}
    \sigma(r)=ce^{-dr^2},
\end{equation}
where $c$ and $d$ are constants. Focusing on a particular case will give us a practical understanding of the interaction of Chern-Simons vortices with magnetic impurities.

\section{Static solutions}
\label{static}

Before moving on to the results, we opt to rescale our variables. We set $x\to 2^{-1}\kappa\eta^{-2} x$, $\phi\to\eta\phi$, $c\to\eta^2c$, and $A_\mu\to2\kappa^{-1}\eta^2A_\mu$. This rescaling effectively amounts to setting $\kappa=2$ and $\eta=1$. Our focus lies on how the system responds to the magnetic impurity. Therefore, we will evaluate the field configurations for different values of $c$, the impurity height, and $d$, the impurity width. Many of the numerical techniques we will use will be adapted from Refs.~\cite{Cockburn:2017, Ashcroft:2020}.

We can solve eqs.~\eqref{eq:BPS1} and \eqref{eq:BPS2} for rotationally symmetric configurations. First, we consider the following ansatz 
\begin{align}
\label{eq:ansatzphi}
    \phi&=g(r)e^{in\theta},\\
    A_i&=\epsilon_{ij}\frac{x^j}{r^2}[n-a(r)],
\label{eq:ansatzA}
\end{align}
where $n$ corresponds to the vortex winding number. Substituting in the BPS equation we find
\begin{align}
    g^\prime&=\frac{ag}{r},\\
    \frac{a^\prime}{r}&=\frac{1}{2}g^2(g^2-1-\sigma).
\end{align} 
These equations describe solutions with spherical symmetry.

In a more general scenario, we can perform the same trick by Taubes \cite{Jaffe:1980} with the difference that we are considering a Chern-Simons instead of a Maxwell term. Defining $u=\log|\phi|^2$, it yields
\begin{equation}
\label{eq:taubes}
    \nabla^2u=e^u(e^u-1-\sigma)+4\pi\sum_{r=1}^N\delta^2(z-z_r),
\end{equation}
where $z=x+iy$ is the spatial coordinate and $z_r$ are the zeros of $\phi$ in the complex plane. Using the equations above, we will compute the vortex profiles for the static configurations. We will focus on the vacuum and the topological vortices.

\subsection{Vacuum Solution}

We will commence by deriving static field configurations from the vacuum solution, characterized by the minimal energy state with zero winding number. These solutions are determined by solving equation~\eqref{eq:taubes} while imposing suitable boundary conditions. Given the absence of zeros in the vacuum configuration, the sum over $r$ simplifies to zero. Using a finite difference scheme to implement such a boundary value problem numerically is straightforward.

\begin{figure}
    \centering
    \subfloat{
      \includegraphics[width=0.48\textwidth]{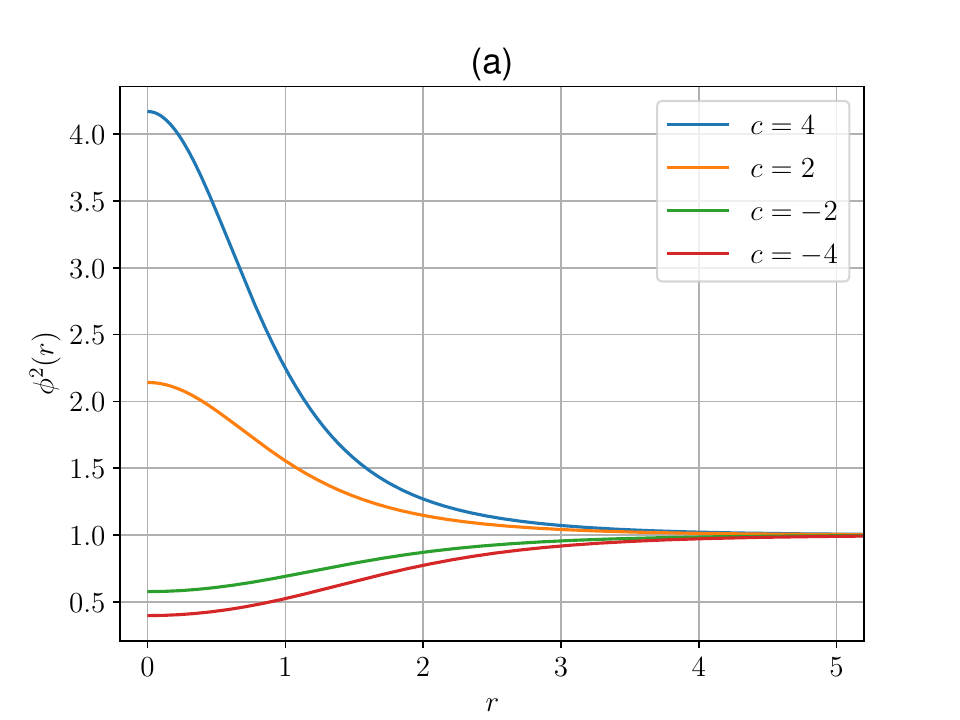}
    }
    \subfloat{
      \includegraphics[width=0.48\textwidth]{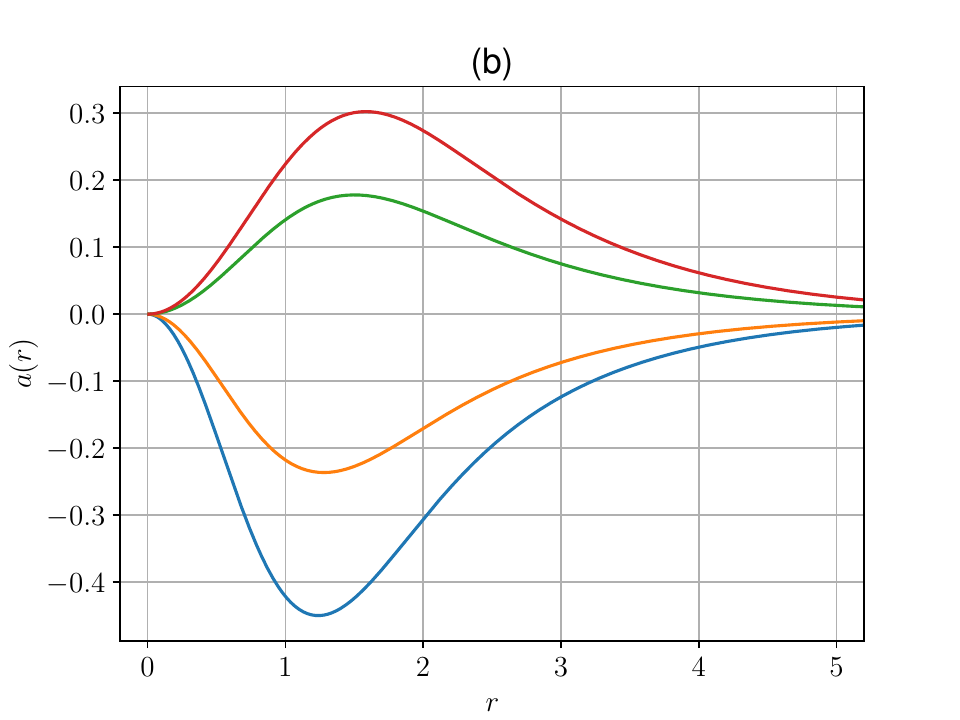}
    }\\
    \subfloat{
      \includegraphics[width=0.48\textwidth]{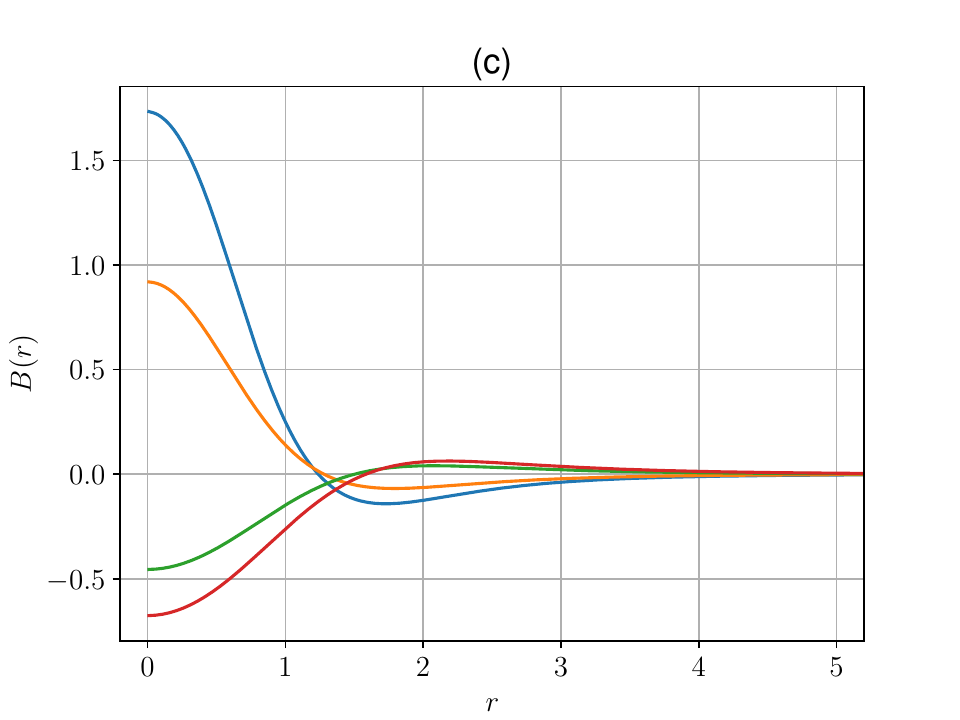}
    }
    \subfloat{
      \includegraphics[width=0.48\textwidth]{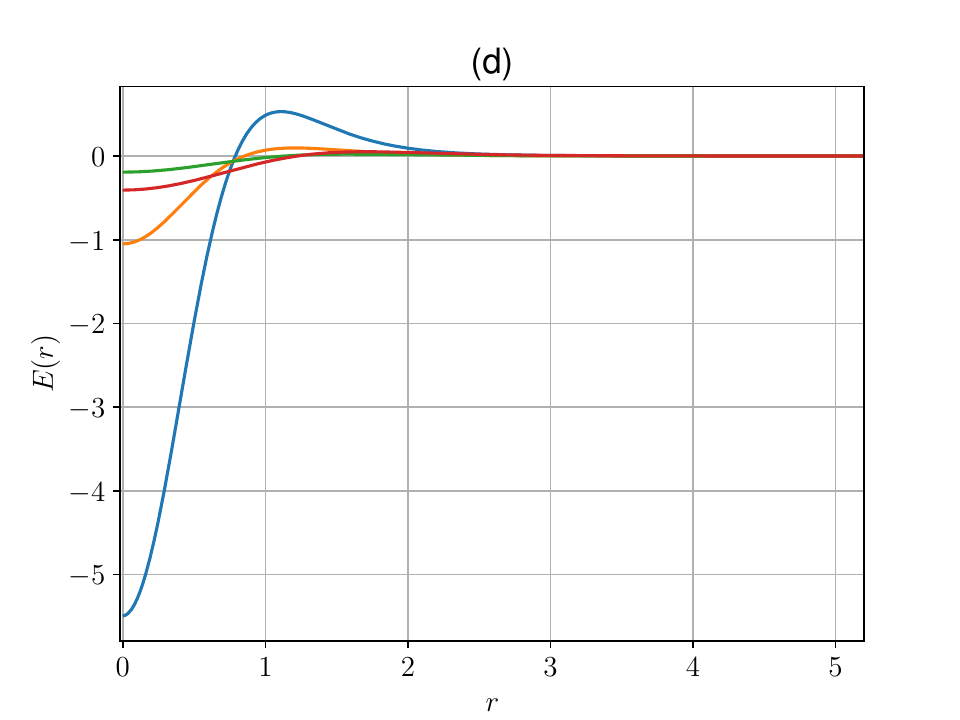}
    }
    \caption{Vacuum profile in the presence of a magnetic impurity as a function of $r$. The panels describe (a) the scalar field, (b) the vector field, (c) the magnetic field, and (d) the energy density. We fixed $d=1$.}
    \label{fig:vacuum}
\end{figure}

The vacuum profile exhibits rotational symmetry, denoted as $\phi=\phi(r)$, owing to its zero winding number. Furthermore, the boundary conditions for our problem can be specified as follows: At the origin, $\phi'(0)=0$ to ensure smoothness. As $r$ tends to infinity, the field approaches the trivial vacuum, $\phi(r\to\infty)=1$, which, in terms of $u$, corresponds to $u'(0)=0$ and $u(r\to\infty)=0$. However, the singularity at $r=0$ demands an analytical Taylor expansion to approximate $\phi(r=\epsilon)$ for a small $\epsilon$. Subsequently, the boundary value algorithm is executed within the interval $[\epsilon, L]$, where the upper limit is set to $L=100.0$. The numerical results are depicted in Fig.~\ref{fig:vacuum}.

The component $\phi$ exhibits nontrivial behavior in the vacuum state due to the impurity, with values greater than one for positive $c$ and less than one for negative $c$. Similarly, the component $a$ is nontrivial as well. Its modulus starts from zero at the origin, progressively increases with radial distance $r$, and eventually diminishes for large $r$. Note that the signs of $a$ and $c$ are opposite. Consequently, the magnetic field $B$ shares the same sign as $c$ at the origin, changes direction as $r$ increases, and ultimately, tapers off for large $r$.
Likewise, the energy density $E(r)$ undergoes sign reversal with increasing $r$ to yield a net energy of zero. Both the integrals of $E(r)$ and $B$ are directly proportional to the winding number $n$, and thus, for the current scenario where $n=0$, they equate to zero.

\subsection{Topological solutions}

\begin{figure}
    \centering
    \subfloat{
      \includegraphics[width=0.48\textwidth]{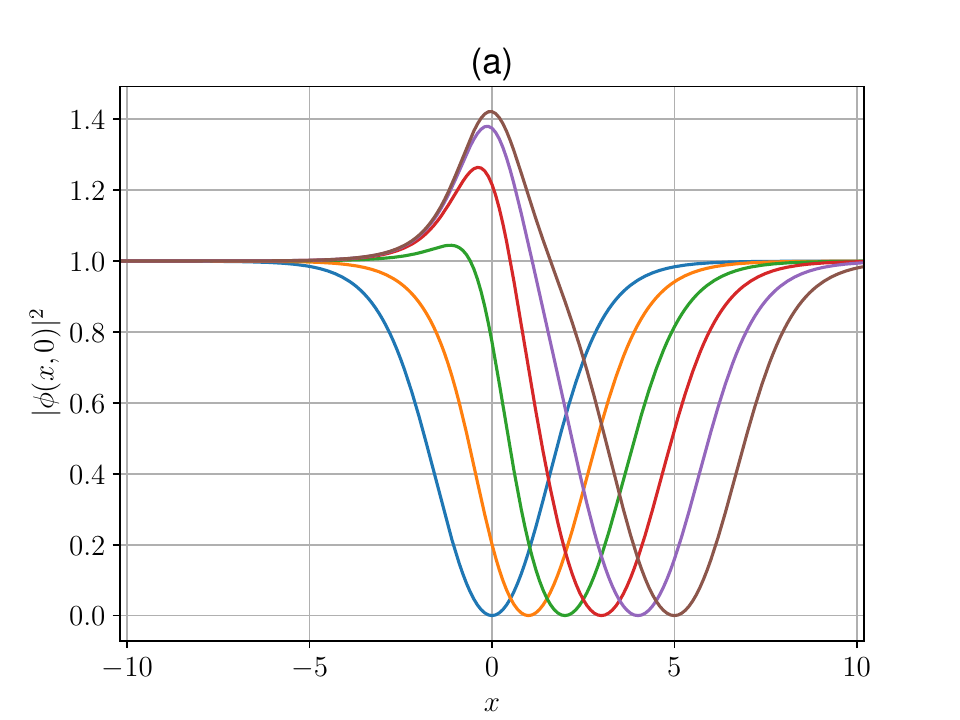}
    }
    \subfloat{
      \includegraphics[width=0.48\textwidth]{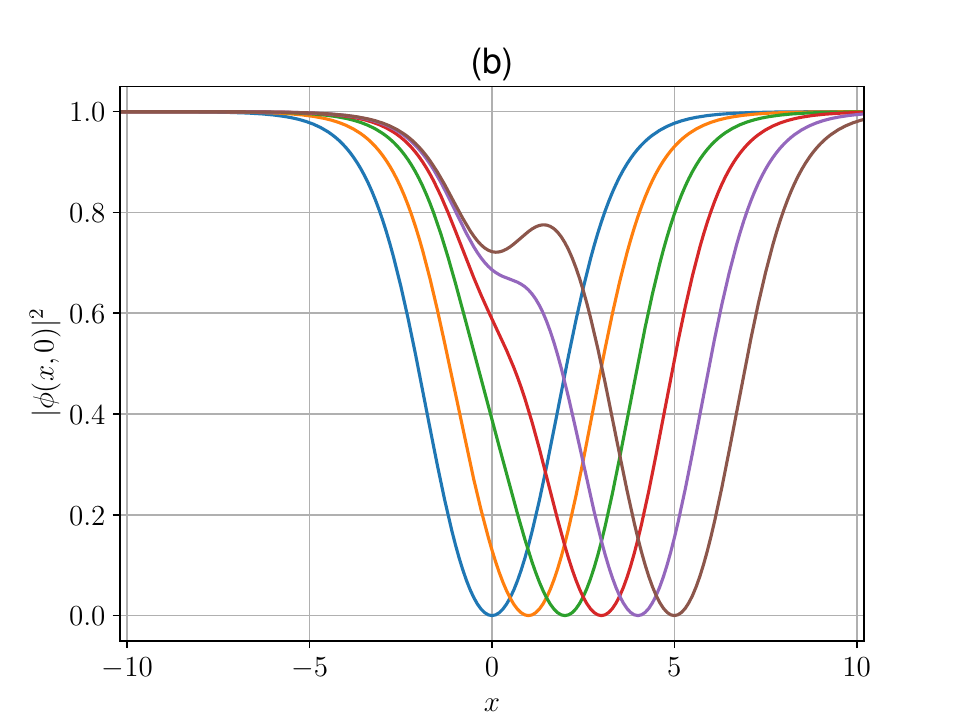}
    }\\
    \caption{A slice of the vortex scalar field profile in the presence of a magnetic impurity. They are centered at positions $z_1=x_1+iy_1$ with $y_1=0$ and $x_1\in\{1,2,3,4,5\}$. We fix $d=1.0$ and the other parameters are (a) $c=1.0$ and (b) $c=-1.0$.}
    \label{fig:top-vortex}
\end{figure}

The next step is to compute the $n=1$ vortex solutions. Here, we employ a different scheme adapted from Ref.~\cite{Cockburn:2017}. First, we dispose of the Dirac delta functions in eq.~\eqref{eq:taubes} by the following change of variables
\begin{equation}
    \mathcal{U}=u-\sum_{r=1}^N\log|z-z_r|^2.
\end{equation}
Then, eq.~\eqref{eq:taubes} becomes
\begin{equation}
    \nabla^2\mathcal{U}=e^\mathcal{U}\prod_{r=1}^{N}|z-z_r|^2\left(e^\mathcal{U}\prod_{r=1}^{N}|z-z_r|^2-1-\sigma\right),
\end{equation}
where $N$ gives the vortex number. Again, we enforce that $u(r\to\infty)=0$ at the boundaries. Formally, we write 
\begin{equation}
\label{eq:bnd}
    \mathcal{U}\sim-\sum_{r=1}^N\log|z-z_r|^2\quad\text{as}\quad|z|\to\infty.
\end{equation}
Here let us consider the case $N=1$. We apply a numerical scheme to relax an initial guess into the correct solution while keeping the boundaries fixed. A suitable initial guess for $\mathcal{U}$, concerning the properties described above, is \cite{Cockburn:2017}
\begin{equation}
    \mathcal{U}=\log\left[\tanh^2\left(\lambda|z-z_1|\right)\right]-\log|z-z_1|^2+\mathcal{U}_{\text{vac}}.
\end{equation}
The first term is an approximation for the $u(z)$ profile of a vortex. We found good results for $\lambda=0.6$. The second term is included to ensure the correct asymptotic behavior of $\mathcal{U}$ (cf. eq.~\eqref{eq:bnd}). Finally, $\mathcal{U}_{\text{vac}}$ is the vacuum solution obtained above. We show the result, after the minimization, in Fig.~\ref{fig:top-vortex}.

The vortex solutions exhibit a fascinating feature. When the vortex center $z_1$ is distant from the origin, the solution closely resembles the isolated vacuum configuration and an isolated vortex of the original theory, i.e., in the absence of a magnetic impurity. Such property justifies the use of the Abrikosov ansatz in the following section. However, when the vortex approaches the impurity, it undergoes partial screening, leading to the disappearance of the bumps and valleys created by the impurity. This phenomenon is a manifestation of the Kondo effect \cite{Kondo:1964}. Essentially, it means that the vortex charge density increases or decreases to cancel the impurity's magnetic field. 

\section{Scattering of vortices by magnetic impurities}

\label{scattering}

\subsection{Moduli space dynamics}

Before computing the vortex evolution via the full equations of motion, let us explore the structure of the moduli space metric for a single vortex. In this section, we aim to develop intuition, which will be helpful when interpreting the scattering output.

To highlight the time-dependent part of the Lagrangian density, we rewrite it as
\begin{equation}
\label{eq:eflag1}
 \mathcal{L}=(\partial_t|\phi|)^2+2B\partial_t\chi-\epsilon^{ij}A_i\partial_tA_j-\mathcal{E}_{\text{static}},
\end{equation}
defining $\phi=|\phi|e^{i\chi}$. For near BPS solutions, the static part of the energy density $\mathcal{E}_{\text{static}}$ is fixed in a given topological sector when integrated. Therefore, the first three terms govern the moduli space dynamics and the effective Lagrangian is given by the integral of these terms over all space.

 First, we know that far from the magnetic impurity, the moduli space is flat because the impurity does not influence the vortex, and it moves freely. It yields the following effective Lagrangian
\begin{equation}
  L_{\text{eff}}=\frac{M}{2}\left(\dot{R}^2+R^2\dot{\Phi}^2\right),\quad\text{for}\quad R\gg d,
\end{equation}
in polar coordinates ($R$,$\Phi$) where $M=2\pi$ is the vortex mass and the dot denotes a time derivative. Notice that the asymptotic behavior for large $R$ contrasts with a two-vortex interaction, which yields an asymptotically conic metric with deficit angle $\pi$. In fact, the vortex-impurity interaction results in an asymptotically flat metric with no deficit angle.

In the opposing limit, $R\ll d$, the effective Lagrangian was found in Ref.~\cite{dziarmaga:1994}. It was obtained by exploring the spherical symmetry of the configuration with $R=0$. Adapting the derivation, we obtain the following expression
\begin{equation}
\label{eq:metric_ap}
  L_{\text{eff}}=\frac{M^\prime}{2}(\dot{R}^2+R^2\dot{\Phi}^2)-QR^2\dot{\Phi},\quad\text{for}\quad R\ll d.
\end{equation}
 The constants $M^\prime$ and $Q$ can be determined through integrals of the field configuration and solutions of the linearized equations of motion around the field configuration.

The first term in the effective Lagrangian \eqref{eq:metric_ap} corresponds to motion in flat space, while the last term represents a magnetic force generated by the impurity's magnetic field. Obtaining the metric in the general case is considerably more involved \cite{dziarmaga:1995}, but not more enlightening. The interaction between the vortex electric charge and the impurity's magnetic field, expressed in the last term in the above Lagrangian, results in markedly distinct scattering outcomes for the ACS model compared to the AMH.

%\begin{equation}
%  \label{eq:metric} L_{\text{eff}}=I(R)\dot{R}^2+2J(R)R\dot{R}\dot{\Phi}+K(R)R^2\dot{\Phi}^2-S(R)R\dot{\Phi}.
%\end{equation}
%The first three terms describe the most general metric with circular symmetry. In particular, the second term breaks chiral symmetry and is present generally in a circularly symmetric metric. Finally, the last term describes magnetic-like force.

\subsection{Equations of motion integration}

One of the key contributions of our study is the development and application of an algorithm for simulating the evolution of ACS vortices. To the best of our knowledge, the only existing implementation of such systems in the literature is found in Ref.~\cite{Strilka:2012}. There, the author considered a different scenario, which was the non-relativistic scattering of two Chern-Simons vortices. Here, our algorithm for scattering the ACS vortex with the magnetic BPS impurity is entirely relativistic.

Our goal is to study the scattering of vortices by the magnetic impurity. The outline of our approach is as follows. The equations of motion that we aim to integrate numerically are the following:
\begin{align}
    \label{eq:2nd}
    D_\mu D^\mu\phi&=-(1+\sigma-3|\phi|^2)(1+\sigma-|\phi|^2)\phi,\\
    \label{eq:gauss}
    -2B&=J^0,\\
    \partial_tA_1&=\partial_xA_0+\frac{1}{2}\left(J^2-\partial_x\sigma\right),\\
    \partial_tA_2&=\partial_yA_0-\frac{1}{2}\left(J^1+\partial_y\sigma\right).
\end{align}
For the Chern-Simons model, eq.~\eqref{eq:gauss} corresponds to Gauss's law. It is a constraint that is held fixed by the field evolution if enforced at the initial condition. Therefore, we need the gauge fixing condition $\partial^\mu A_\mu=0$, to compute the evolution of $A_0$, that is
\begin{equation}
  \partial_tA_0=\partial_iA_i.
\end{equation}
Then, we split the second-order equation \eqref{eq:2nd} into two equations. As the conjugate momentum is given by $\pi=D_0\phi=\partial_t\phi+iA_0\phi$, it leads to
\begin{align}
    \partial_t\pi&=D_iD_i\phi-(1+\sigma-3|\phi|^2)(1+\sigma-|\phi|^2)\phi-iA_0\pi\\
    \partial_t\phi&=\pi-iA_0\phi.
\end{align}
We must also select suitable boundary conditions. We opt for $D_i\phi=0$ and $\partial_1A_1=\partial_2A_2=0$. Since our focus lies solely on bulk evolution, i.e., the evolution far from the boundary, our findings do not depend on this choice.

The numerical method we employ involves discretizing spatial derivatives within a box spanning the interval $[-16,16]\times[-16,16]$, utilizing a five-point stencil approximation. Specifically, for any field $\psi$, these conditions read
\begin{equation}
    \partial_i\psi=\big[-\psi(\boldsymbol{x}+2h\boldsymbol{e}_i)+8\psi(\boldsymbol{x}+h\boldsymbol{e}_i)-8\psi(\boldsymbol{x}-h\boldsymbol{e}_i)+\psi(\boldsymbol{x}-2h\boldsymbol{e}_i)\big]/12h+\mathcal{O}(h^5),
\end{equation}
and
\begin{equation}
    \partial_i^2\psi=\big[-\psi(\boldsymbol{x}+2h\boldsymbol{e}_i)+16\psi(\boldsymbol{x}+h\boldsymbol{e}_i)-30\psi(\boldsymbol{x})+16\psi(\boldsymbol{x}-h\boldsymbol{e}_i)-\psi(\boldsymbol{x}-2h\boldsymbol{e}_i)\big]/12h^2+\mathcal{O}(h^5).
\end{equation}

At the boundaries, we reduce the order of derivative approximations to ensure compatibility with the finite box size. The resulting differential equations are integrated using a third-order Runge-Kutta algorithm with adaptive step size and error control, implemented through the SciPy library in Python.

The numerical error is estimated by the deviation of the total charge, magnetic flux, and energy of the field configuration from their initial values. Moreover, we verify Gauss's law by comparing the magnetic field and charge density at every point, as well as the total value of the magnetic flux and the electric charge. Any simulations with a relative error exceeding one percent in any test were discarded and repeated with a smaller grid spacing $h$. Typically, we considered $h=0.125$ or $0.0625$.

In addition to the equations of motion, we also require the initial condition. To begin, we construct a boosted vortex solution. Starting with the vortex solutions $\phi$ and $A^\mu$ in the absence of the impurity, the boosted solutions are obtained via
\begin{align}
\phi^\prime(x,y,t)&=\phi(\gamma(x-vt),y,\gamma(t-vx)),\\
A^{\mu^\prime}(x,y,t)&={\Lambda^{\mu^\prime}}_\nu A^\nu(\gamma(x-vt),y,\gamma(t-vx)),
\end{align}
where the boost is in the positive $x$-direction and ${\Lambda^{\mu^\prime}}_\nu$ is its corresponding Lorentz transformation matrix. 

The next step is to construct vortex-impurity solutions via the Abrikosov ansatz, which reads
\begin{equation}
    \hat{\phi}=\prod_{i=1}^2\phi^{(i)}(x-x_i),\quad\hat{A}=\sum_{i=1}^2A_\mu^{(i)}(x-x_i),
\end{equation}
where $\phi^{(i)}$ and $A_\mu^{(i)}$ are either the impurity solution ($i=I$) centered at the origin or the boosted vortex solution ($i=V$) centered at $x_V$.

\subsection{Vortex with winding number one}

Let us first analyze the collision between a vortex with winding number $n=1$ and a magnetic impurity. The impurity is initially at the origin, while the vortex is located at $\boldsymbol{x}_V=(-8,b)$, where $b$ represents the impact parameter.

\begin{figure}
    \centering
      \includegraphics[width=0.96\textwidth]{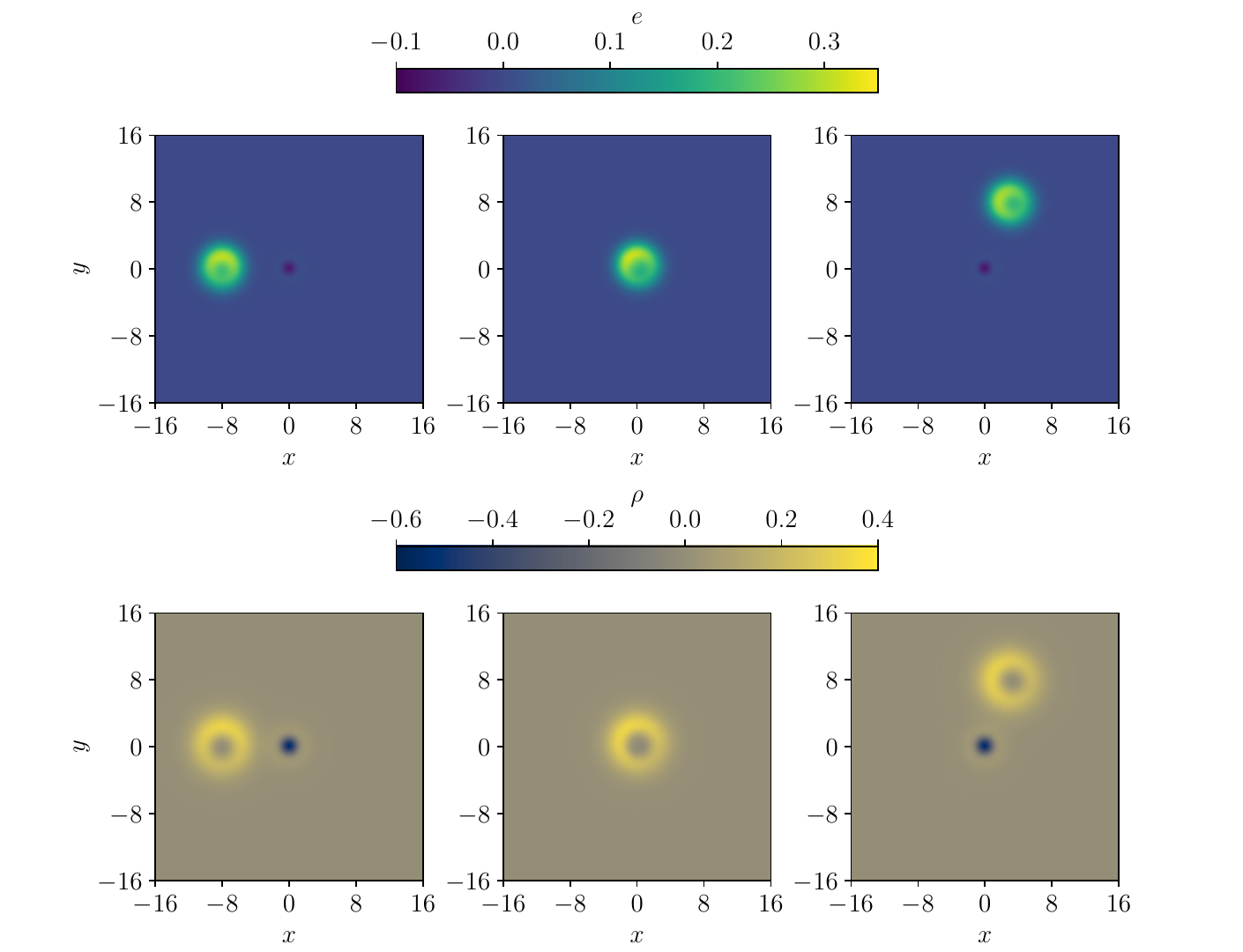}
    \caption{Snapshots of the energy and charge densities, $e$ and $\rho$, for a vortex-impurity collision. Parameters are $v=0.2$, $(c,d)=(-1.0,1.0)$, $L=16.0$, and $(x_0,b)=(8.0,0.0)$. The instants of time are (left) $t=0.0$, (middle) $t=44.0$, and (right) $t=88.0$. }
    \label{fig:col}
\end{figure}

In Fig.~\ref{fig:col}, we show three snapshots for a collision with parameters $v=0.2$, $b=0.0$, $c=-1.0$, and $d=1.0$. The vortex approaches the impurity, reaches the origin, and then recedes from the impurity at a positive angle (anti-clockwise). Notice that the impurity is partially obscured with the vortex near the origin, in agreement with Fig.~\ref{fig:top-vortex}.

As shown in Ref.~\cite{Cockburn:2017}, AMH vortices are not deflected by a magnetic impurity at zero impact parameter. However, the figure shows that ACS vortices are deflected in the same scenario. The discrepancy is explained by the following arguments. It was shown in Refs.~\cite{dziarmaga:1994, dziarmaga:1995} that the moduli space of ACS vortices exhibits a metric structure closely resembling that of the AMH model in the following way. It has polar symmetry because the impurity depends only on the radius. Moreover, the moduli spaces tend to $R^2$ asymptotically for large $r$ due to the impurity's diminishing effect. Such a metric does not deflect the particle for a radial motion. However, the effective Lagrangian described by equation \eqref{eq:metric_ap} introduces an additional term proportional to $R^2\dot{\Phi}$, describing a magnetic-like force. It is the main driver of the observed deflection, as we will see below. Moreover, if it were absent, the vortex deflection observed in Fig.~\ref{fig:col} would not occur. 

\begin{figure}
    \centering
    \subfloat{
      \includegraphics[width=0.48\textwidth]{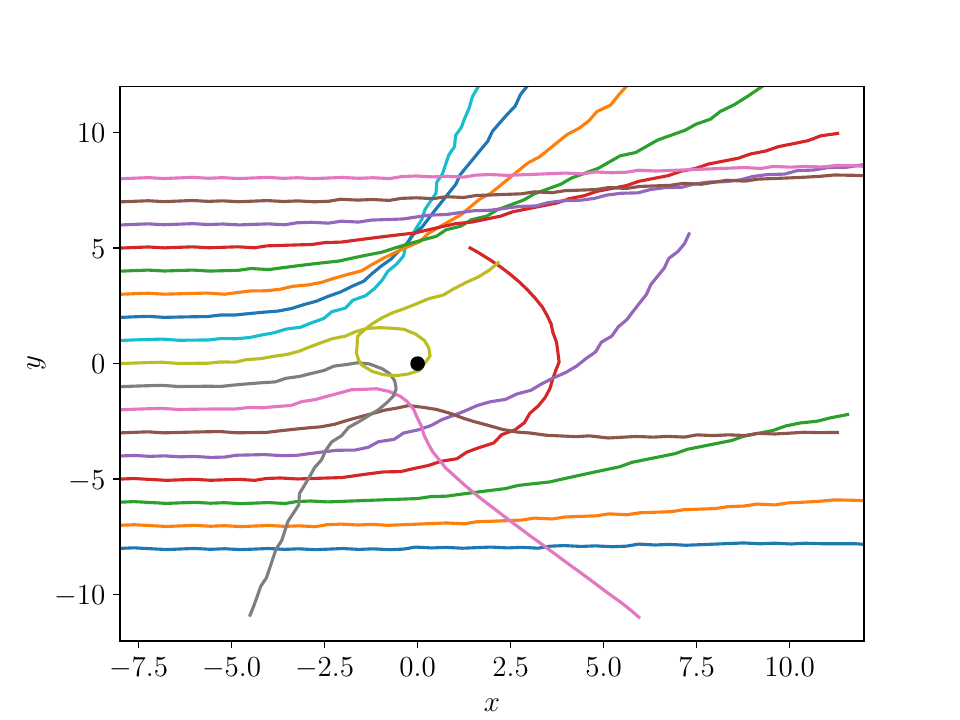}
    }
    \subfloat{
      \includegraphics[width=0.48\textwidth]{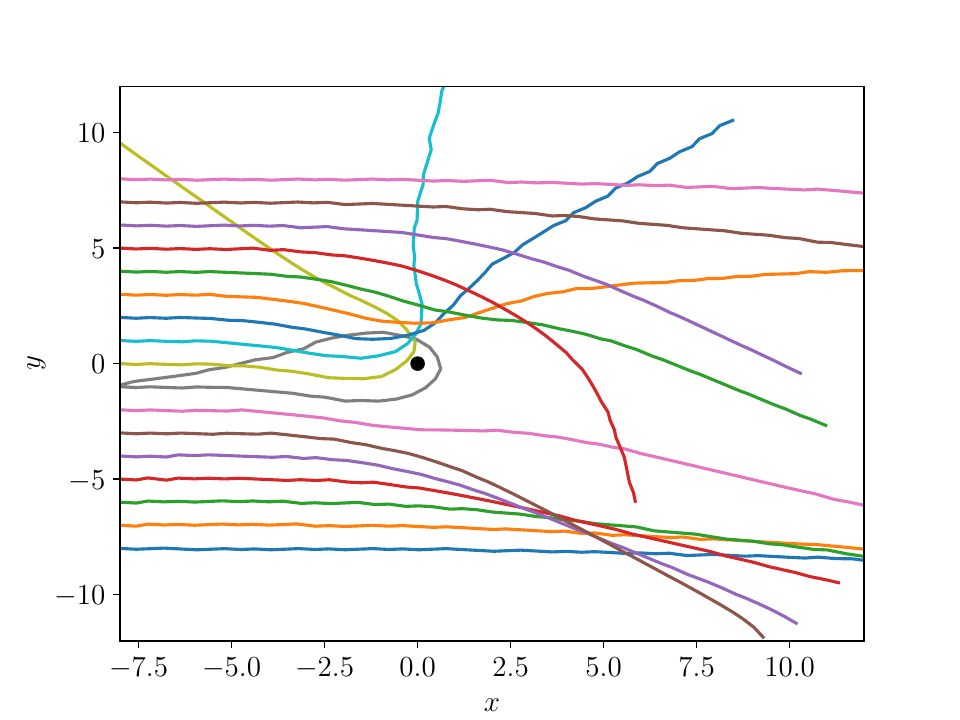}
    }\\
    \subfloat{
      \includegraphics[width=0.48\textwidth]{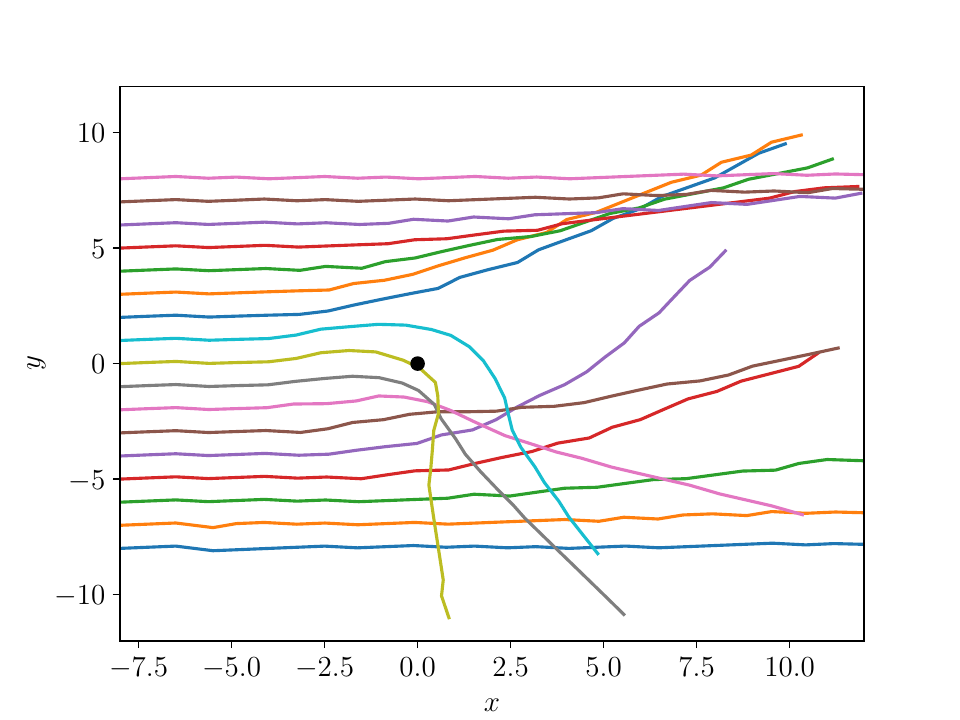}
    }
    \subfloat{
      \includegraphics[width=0.48\textwidth]{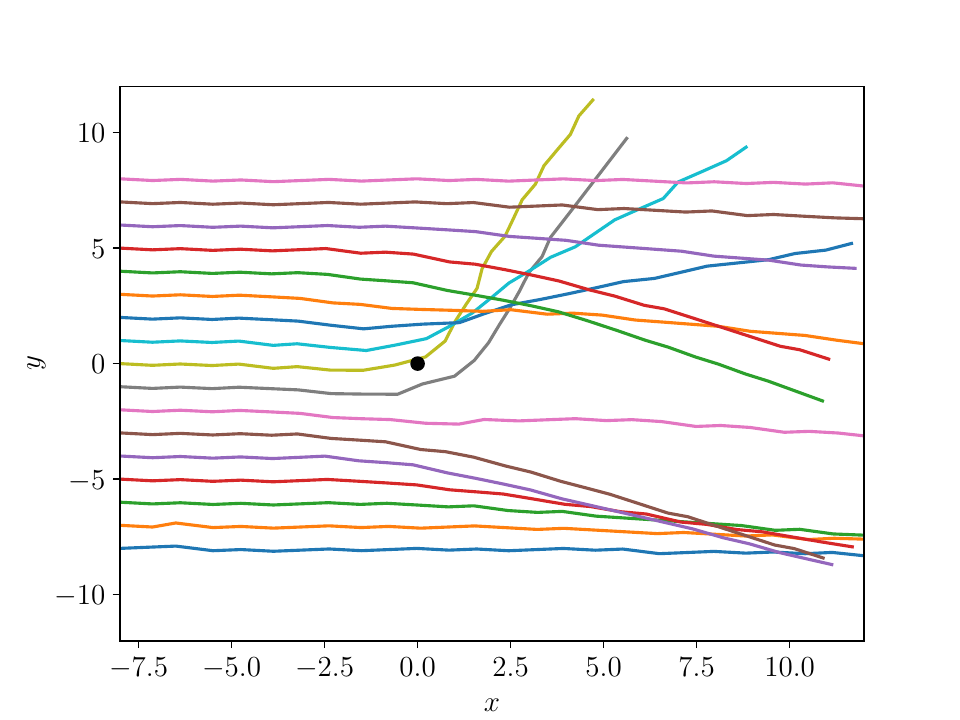}
    }\\
    \subfloat{
      \includegraphics[width=0.48\textwidth]{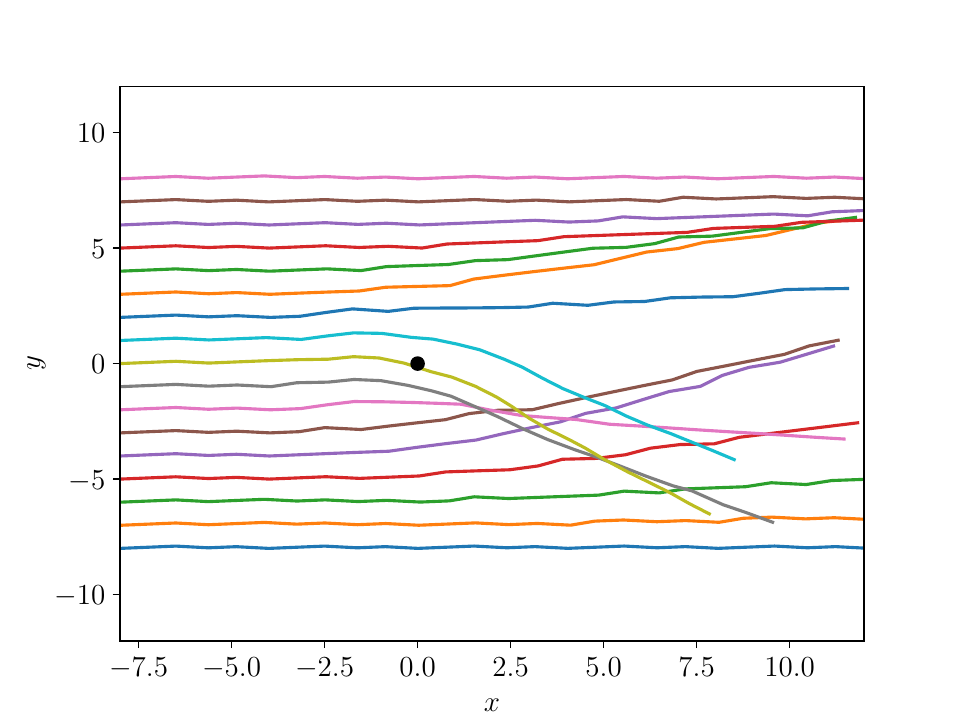}
    }
    \subfloat{
      \includegraphics[width=0.48\textwidth]{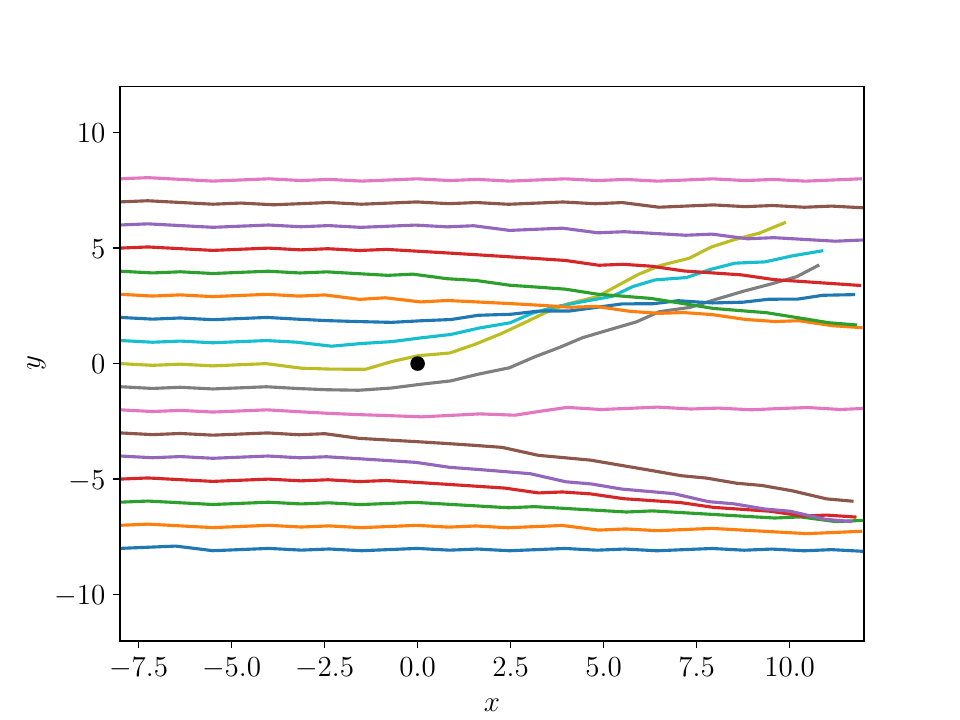}
    }\\
    \caption{Vortex path during a collision with an impurity for several impact parameters. Parameters are $c=1.0$ (left) and $c=-1.0$ (right) and $v=0.1$ (top), $v=0.2$ (middle), and $v=0.4$ (bottom). The black dot locates the impurity.}
    \label{fig:path}
\end{figure}

In Fig.~\ref{fig:path}, we illustrate several vortex trajectories corresponding to different values of $v$, $b$, and $c$. The vortex center is computed numerically in the following manner. Let us define a plaquette $P_{(x,y)}$ as the set of points given by
\begin{equation}
    P_{(x,y)}=\{(x,y),(x+h,y),(x,y+h),(x+h,y+h)\}.
\end{equation}
The vortex position is defined as the point where $\phi=0$. Now, notice that the vortex field has the following property
\begin{equation}
    \oint \phi(z)dz=2N\pi,
\end{equation}
where $N$ is the number of enclosed vortices. Thus, we search for a plaquette where the phase of $\phi$ changes by $2\pi$, meaning that it contains a vortex. Afterward, we find a polynomial interpolation for $\phi$ at that plaquette. Then, the vortex position is estimated by the root $z_r=x_r+iy_r$ of the polynomial inside that region. In other words, the $z_r$ is our best estimate of the position where $\phi$ vanishes, i.e., the vortex position.

The analysis of vortex paths reveals several notable features. First, there is a tendency for deflection to decrease as velocity increases, which is compatible with the expected behavior of charge deflection in a magnetic field, where the radius of orbit expands with velocity. Second, as the impact parameter varies from -8 to 8, the deflection angle undergoes two sign changes (see also Fig.~\ref{fig:deflexion} below). This is attributed to the magnetic field's profile depicted in Fig.~\ref{fig:vacuum}, which changes sign as the distance from the origin increases. Third, the deflection angle changes sign primarily by reversing the sign of $c$ due to the magnetic field's inversion at the origin. Fourth, the asymmetry between positive and negative $b$ arises from the chiral nature of ACS theory. Moreover, the asymmetry between positive and negative $c$ stems from the asymmetric impact of positive and negative impurities on the vortices, as illustrated in Fig. \ref{fig:top-vortex}.

The vortex scattering output exhibited in Fig.~\ref{fig:path} is succinctly summarized in Fig.~\ref{fig:deflexion}, where we illustrate the deflection angle $\Theta$ of the vortices as a function of the impact parameter $b$. As expected from a magnetic force, $\Theta$ diminishes with increasing velocity. Dziarmaga's conjecture \cite{dziarmaga:1994} suggests that at high velocities, the magnetic deflection diminishes due to the expanding orbit radius, leaving only the metric component of the effective Lagrangian with two time derivatives. Here, we observe a decreasing deflection with $v$, which is in agreement with a larger orbit radius for the magnetic force. However, the full conjecture is complex to test because the moduli space approximation eventually breaks down as the velocity increases. After all, the system moves away from the BPS limit.

\begin{figure}
    \centering
    \subfloat{
      \includegraphics[width=0.48\textwidth]{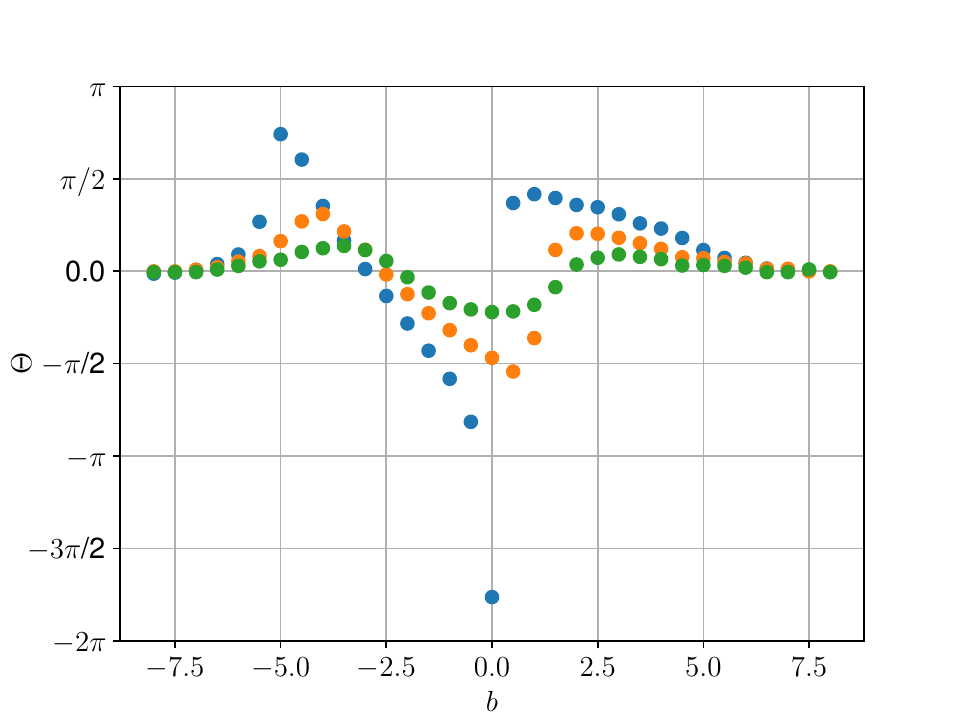}
    }
    \subfloat{
      \includegraphics[width=0.48\textwidth]{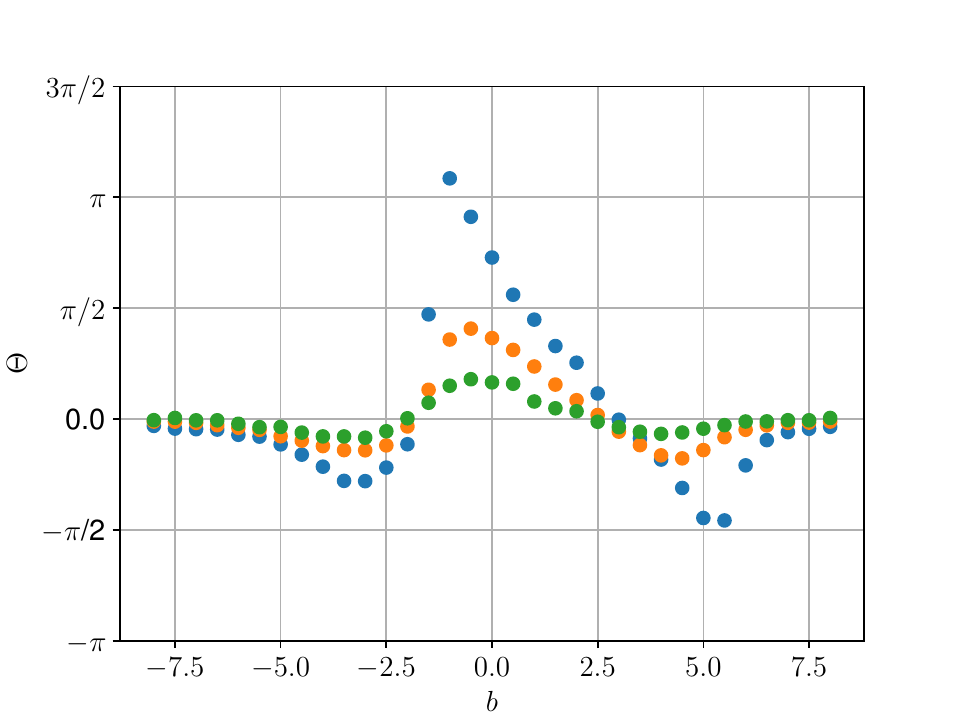}
    }\\
    \caption{Deflection angle for a vortex-impurity collision as a function of the impact parameter. The panels correspond to (left) $c=1.0$ and (right) $c=-1.0$. The colors correspond to (blue) $v=0.1$, (orange) $v=0.2$, and (green) $v=0.4$.}
    \label{fig:deflexion}
\end{figure}

\subsection{Vortex with winding number two}

\begin{figure}
    \centering
    \subfloat{
      \includegraphics[width=0.96\textwidth]{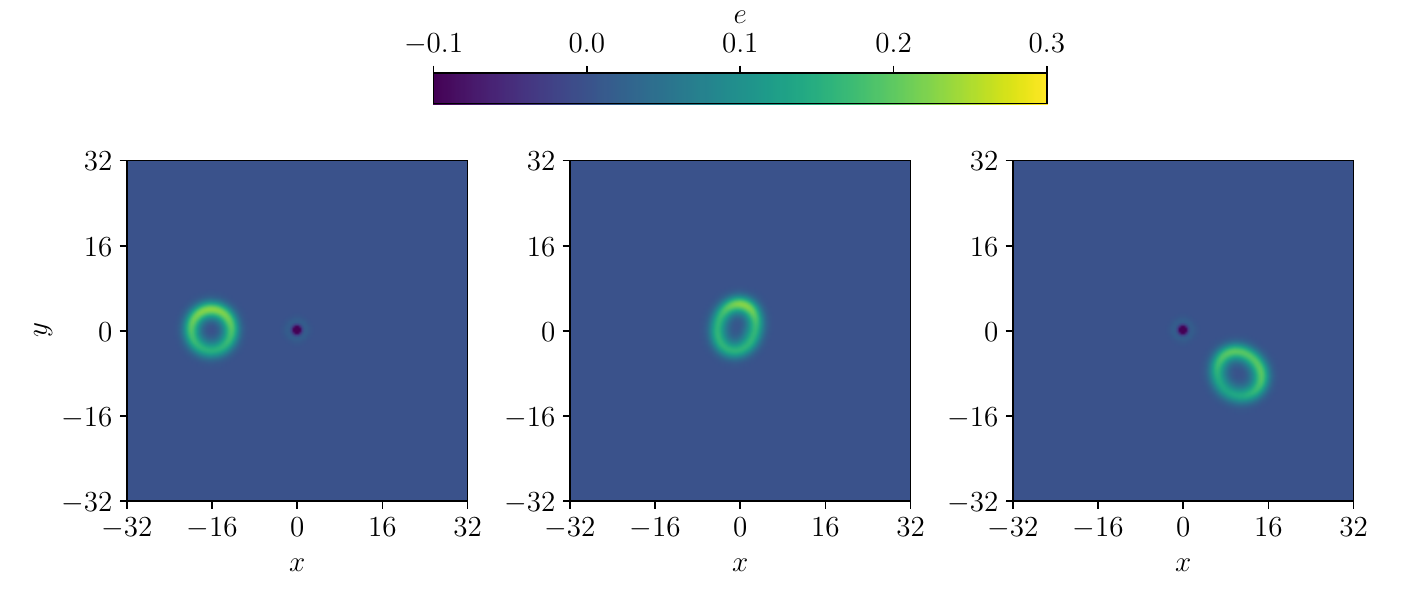}
    }\\
    \subfloat{
      \includegraphics[width=0.96\textwidth,trim={0 0 0 2.5cm},clip]{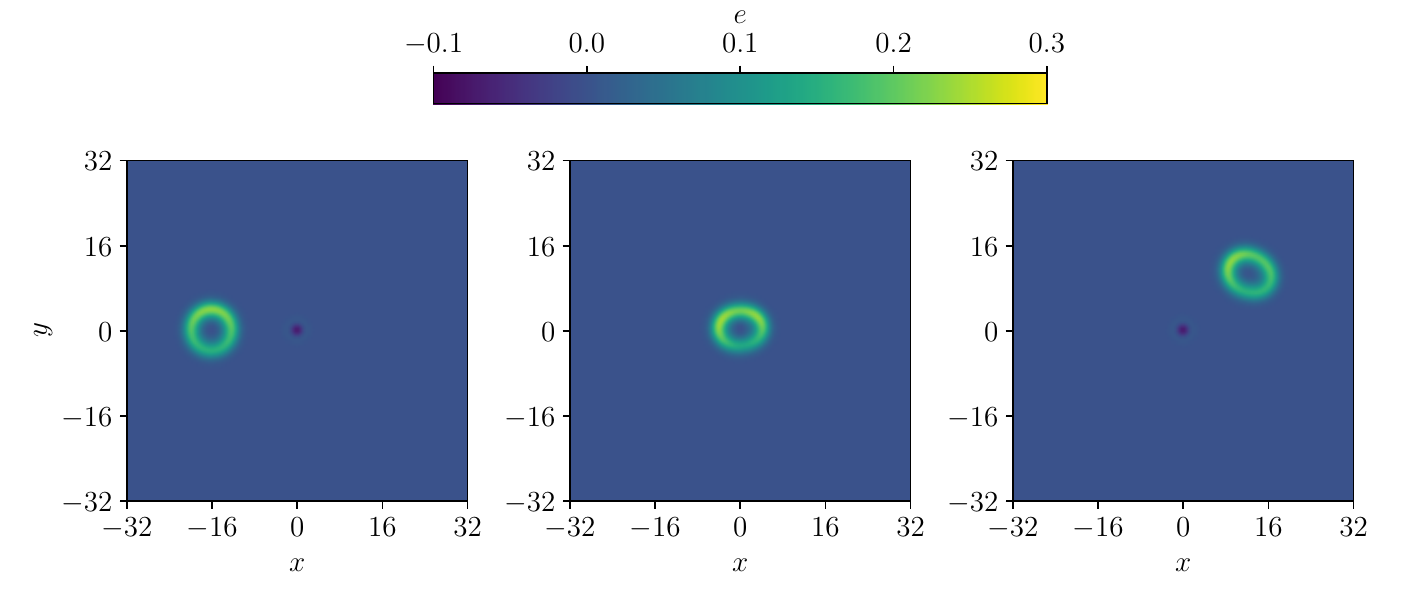}
    }\\
    \caption{Snapshots of the energy density during a collision between vortices with winding number two and a magnetic impurity. In the upper panel, $c=1.0$ and snapshots are taken at $t=0.0$, $t=84.0$, and $t=168.0$. In the lower one, $c=-1.0$ and snapshots are taken at $t=0.0$, $t=80.0$, and $t=160.0$. Other parameters are $d=0.5$, $v=0.2$, $L=32.0$, and $(x_0,b)=(16.0,0.0)$. }
    \label{fig:coln2}
\end{figure}

We now examine the behavior of vortices with a winding number two when scattered by a magnetic impurity. Initially, we consider a single spherically symmetric vortex with $n=2$. Loosely speaking, we are considering two vortices positioned atop each other. We utilize a larger Gaussian impurity with $d=0.5$ and expand the computational domain to $[-32,32]\times[-32,32]$ to accommodate the larger vortex size. The vortex is boosted on the positive $x$-direction, and we set the initial position to $\boldsymbol{x}_V=(-16,0)$, with zero impact parameter. The result is shown in Fig.~\ref{fig:coln2}.

The scattering output reveals two distinct effects. First, there is a noticeable magnetic deflection. Moreover, the $n=2$ vortex is stretched after being scattered, acquiring an elliptical form. For $c=1.0$,  the stretching aligns with the velocity direction, while for $c=-1.0$, it occurs perpendicular to the velocity. In essence, this implies that the ellipse´s major axis is parallel to the velocity for $c=1.0$. On the other hand, the ellipse´s major axis is perpendicular to the velocity for $c=-1.0$. Notably, these arrangements were also observed in the scattering of AMH vortices \cite{Ashcroft:2020}.

\section{Conclusion}

In this work, we analyzed the Abelian Chern-Simons model with a magnetic impurity that preserves half of the system's BPS property. Employing numerical techniques, we computed both its static and scattering properties. The static solution of an isolated magnetic impurity exhibits either a bump or a valley in the scalar field, depending on the direction of the magnetic field. Nevertheless, the magnetic field shifts direction with increasing radial coordinates since the total magnetic flux is zero. 

Solutions with unit winding number were also computed. We found that there exist two regions for most cases, one containing a Chern-Simons vortex and another containing the impurity solution. However, when the two solutions approach each other, the vortex screens the impurity due to Kondo's effect \cite{Kondo:1964}. 

We developed a numerical algorithm to evolve the fields according to the equations of motion, marking the first fully relativistic algorithm reported in the literature to the best of our knowledge. Our investigation revealed significant differences in the scattering outcomes between an ACS vortex and a magnetic impurity compared to the AMH model. This discrepancy primarily stems from the presence of magnetic force interaction in the former model, agreeing with the moduli space analysis conducted three decades ago by Dziarmarga \cite{dziarmaga:1994, dziarmaga:1995}, which we adapted for the current scenario.

We also explored the scattering of $n=2$ ACS vortices and a magnetic impurity. It showed that these vortices experience stretching, similar to the AMH model, along also with magnetic deflection. This magnetic deflection stands out as the primary difference between the two models.

An intriguing direction for future exploration involves simulating vortex-impurity scattering for the Schr\"{o}dinger Chern-Simons vortex \cite{Manton:1997, krusch:2006}, which is a great candidate for vortex dynamics in condensed matter superconductors. It would also be interesting to use the numerical method described here to simulate collisions involving more than two vortices and also a vortex and an anti-vortex, complementing the findings documented in Ref.~\cite{Strilka:2012}.

An essential challenge lies in obtaining a numerical estimate for all functions within the moduli space effective Lagrangian. In Ref.~\cite{Kim:1994}, the Lagrangian was found in terms of the field configuration and linearized perturbations around the field configurations. Then, two approaches were developed. In Ref.\cite{dziarmaga:1995}, the author employed an adiabatic approximation by assuming that the linearized perturbations are linear in velocity. However, this resulted in a highly complex system of equations describing the Lagrangian, making it difficult to solve, even numerically, due to its nonlinear nature. In Ref.~\cite{Nesterov:1998}, on the other hand, took a crude approach by neglecting all linearized perturbations, leading to inaccuracies such as an incorrect estimation of the mass of a single vortex in the effective Lagrangian. Therefore, it is crucial to either solve the full equations in Ref.~\cite{dziarmaga:1995} or at least find a better approximation than the one in Ref.~\cite{Nesterov:1998}. This is a problem currently under investigation.

\section*{Acknowledgements}

DB would like to thank the National Council for Scientific and Technological Development (CNPq) Grants no. 303469/2019-6 and no. 402830/2023-7, and Paraiba State Research Foundation, Grant no. 0015/2019. JGFC acknowledges financial support from Fundação de Amparo a Ciência e Tecnologia do Estado de Pernambuco (FACEPE), grant no. BFP-0013-1.05/23. AM acknowledges financial support from  CNPq, Grant no. 309368/2020-0, and Coordenação de Aperfeiçoamento de Pessoal de Nível Superior (CAPES). Part of the simulations presented here were performed in the supercomputer SDumont at the Brazilian agency LNCC (Laboratório Nacional de Computação Científica).


\begin{thebibliography}{99}

\bibitem{Landau:2013} L. D. Landau and E. M. Lifshitz, Fluid Mechanics: Landau and Lifshitz: Course of Theoretical Physics, Volume 6 (Vol. 6), Elsevier (2013)

\bibitem{Salomaa:1987} M. M. Salomaa and G. E. Volovik, Quantized vortices in superfluid $^3$He, Rev. Mod. Phys. 59, 533 (1987).

\bibitem{Ginzburg:1950} V.L. Ginzburg and L.D. Landau, On the Theory of superconductivity, Zh. Eksp. Teor. Fiz. 20, 1064 (1950).

\bibitem{Abrikosov:1957} A.A. Abrikosov, On the Magnetic properties of superconductor of the second group, Zh. Eksp. Teor. Fiz. 32, 1442 (1957).

\bibitem{Vilenkin:1994} A. Vilenkin and E. P. S. Shellard, Cosmic strings and other topological defects, Cambridge University Press (1994).

\bibitem{James:1994} M. James, L. Perivolaropoulos, and T. Vachaspati, Stability of electroweak strings, Phys. Rev. D 46, R5232(R) (1994).

\bibitem{Tong:2002} D. Tong, NS5-Branes, T-Duality and Worldsheet Instantons, J. High Energ. Phys. 7, 13 (2002).

\bibitem{Manton:2004} N. Manton and P. Sutcliffe, Topological solitons, Cambridge University Press (2004).

\bibitem{Higgs:1964} P. W. Higgs, Broken Symmetries and the Masses of Gauge Bosons, Phys. Rev. Lett. 13, 508 (1964).

\bibitem{Hong:1990} Jooyoo Hong, Yoonbai Kim, and Pong Youl Pac, Multivortex solutions of the Abelian Chern-Simons-Higgs theory, Phys. Rev. Lett. 64, 2230 (1990).

\bibitem{Jackiw:1990b} R. Jackiw and Erick J. Weinberg, Self-dual Chern-Simons vortices, Phys. Rev. Lett. 64, 2234 (1990).

\bibitem{Jackiw:1990a} R. Jackiw, Kimyeong Lee, and Erick J. Weinberg, Self-dual Chern-Simons solitons, Phys. Rev. D 42, 3488 (1990).

\bibitem{Atland:2010} A. Altland and B. D. Simons, Condensed matter field theory, Cambridge university press (2010).

\bibitem{Harden:1963} J. L. Harden and V. Arp, The lower critical field in the Ginzburg-Landau theory of superconductivity, Cryogenics 3, 105 (1963).

\bibitem{Plohr:1981} B. Plohr, The behavior at infinity of isotropic vortices and monopoles, J. Math. Phys. 22, 2184 (1981).

\bibitem{Berger:1989} M. S. Berger and Y. Y. Chen, Symmetric vortices for the Ginzberg-Landau equations of superconductivity and the nonlinear desingularization phenomenon, J. Func. Anal. 82, 259 (1989).

\bibitem{Jaffe:1980} A. Jaffe and C.H. Taubes, Vortices and Monopoles: Structure of Static Gauge Theories, Birkhäuser Boston (1980).

\bibitem{Bazeia:2018c} D. Bazeia, L. Losano, M. A. Marques, R. Menezes, Analytic vortex solutions in generalized models of the Maxwell–Higgs type, Phys. Lett. B 778, 22 (2018).

\bibitem{Bazeia:2010} D. Bazeia, E. da Hora, R. Menezes, H. P. de Oliveira, and C. dos Santos, Compactlike kinks and vortices in generalized models, Phys. Rev. D 91, 125016 (2010).

\bibitem{Bazeia:2018} D. Bazeia, L. Losano, M. A. Marques, R. Menezes, and I. Zafalan, First order formalism for generalized vortices, Nucl. Phys. B 934, 212 (2018).

\bibitem{Bazeia:2018b} D. Bazeia, M. A. Marques, and R. Menezes, Maxwell–Higgs vortices with internal structure, Phys. Lett. B 780, 485 (2018).

\bibitem{Bazeia:2019} D. Bazeia, M. A. Liao, M. A. Marques, and R. Menezes, Multilayered vortices, Phys. Rev. Res. 1, 033053 (2019).

\bibitem{Andrade:2020} I. Andrade, D. Bazeia, M. A. Marques, and R. Menezes, Long range vortex configurations in generalized models with Maxwell or Chern-Simons dynamics, Phys. Rev. D 102, 025017 (2020).

\bibitem{Ruback:1988} P. J. Ruback, Vortex string motion in the abelian Higgs model, Nucl. Phys. B 296, 669 (1988).

\bibitem{Shellard:1988} E. P. S. Shellard and P. J. Ruback, Vortex scattering in two dimensions, Phys. Lett. B 209, 262 (1988).

\bibitem{Strilka:2012} R. J. Strilka, Low-energy vortex dynamics in the self-dual Chern–Simons–Higgs model, Comm. Nonlinear Sci. Numer. Simulat. 17(10), 3811 (2012).

\bibitem{manton:1982} N. S. Manton, A remark on the scattering of BPS monopoles, Phys. Lett. B 110, 54 (1982).

\bibitem{manton:1985} N. S. Manton, Monopole interactions at long range, Phys. Lett. B 154, 397 (1985).

\bibitem{samols:1992} T. M. Samols, Vortex Scattering, Commun. Math. Phys. 145, 149 (1992).

\bibitem{Strachan:1992} I. A. B. Strachan, Low‐velocity scattering of vortices in a modified Abelian Higgs model, J. Math. Phys. 33, 102 (1992)

\bibitem{dziarmaga:1994} J. Dziarmaga, Short range interactions of Chern-Simons vortices, Phys. Lett. B 320, 3488 (1994).

\bibitem{dziarmaga:1995} J. Dziarmaga, More on scattering of Chern-Simons vortices, Phys. Rev. D 51, 7052 (1995).

\bibitem{Hook:2013} A. Hook, S. Kachru and G. Torroba, Supersymmetric defect models and mirror symmetry, J. High Energ. Phys. 11, 4 (2013).

\bibitem{Tong:2014} D. Tong and K. Wong, Vortices and impurities, J. High Energ. Phys. 1, 90 (2014).

\bibitem{Han:2016} X. Han and Y. Yang, Magnetic impurity inspired Abelian Higgs vortices, J. High Energ. Phys. 2, 46 (2016).

\bibitem{kachru:2010} S. Kachru, A. Karch, and S. Yaidax, Holographic lattices, dimers, and glasses, Phys. Rev. D 81, 026007 (2010).

\bibitem{benicasa:2012a} P. Benincasa and A. V. Ramallo, Fermionic impurities in Chern-Simons-matter theories, J. High Energ. Phys. 2, 76 (2012).

\bibitem{benicasa:2012b} P. Benincasa and A. V. Ramallo, Holographic Kondo model in various dimensions, J. High Energ. Phys. 6, 133 (2012).

\bibitem{Cockburn:2017} Alexander Cockburn, Steffen Krusch, and Abera A. Muhamed, Dynamics of vortices with magnetic impurities, J. Math. Phys. 58, 063509 (2017).

\bibitem{Ashcroft:2020} Jennifer Ashcroft and Steffen Krusch, Vortices and magnetic impurities, Phys. Rev. D 101, 025004 (2020).

\bibitem{Almeida:2022} V. Almeida, R. Casana, E. da Hora, and S. Krusch, Self-dual $CP(2)$ vortex-like solitons in the presence of magnetic impurities, Phys. Rev. D 106, 016010 (2022).

\bibitem{Bazeia:2022} D. Bazeia, M.A. Liao, and M.A. Marques, Impurity-like solutions in vortex systems coupled to a neutral field, Phys. Lett. B 825, 136862 (2022).

\bibitem{Kondo:1964} Jun Kondo, Resistance Minimum in Dilute Magnetic Alloys, Progress of Theoretical Physics 32(1), 37 (1964).

\bibitem{Manton:1997} N. S. Manton, First Order Vortex Dynamics, Annals of Physics 256(1), 114 (1997).

\bibitem{krusch:2006} S. Krusch and P. Sutcliffe, Schrödinger–Chern–Simons vortex dynamics, Nonlinearity 19, 1515 (2006).

\bibitem{Kim:1994} Y. Kim and K. Lee, Vortex Dynamics in Self-Dual Chern-Simons Higgs Systems, Phys. Rev. D 49, 2041 (1994).

\bibitem{Nesterov:1998} A. I. Nesterov, On Vortex Dynamics in the Self-Dual Chern–Simons–Higgs System, Lett. Math. Phys. 45, 203 (1998).

\end{thebibliography}
\end{document}